# Anisotropic mechanical and optical response and negative Poisson's ratio in Mo$_2$C nanomembranes revealed by first-principles simulations

Bohayra Mortazavi[*,1], Masoud Shahrokhi[2], Meysam Makaremi[3], Timon Rabczuk[#,4]

[1]*Institute of Structural Mechanics, Bauhaus-Universität Weimar, Marienstr. 15, D-99423 Weimar, Germany.*

[2]*Institute of Chemical Research of Catalonia, ICIQ, The Barcelona Institute of Science and Technology, Av. Països Catalans 16, ES-43007 Tarragona, Spain.*

[3]*Chemical Engineering Department, Carnegie Mellon University, Pittsburgh, Pennsylvania 15213, USA.*

[4]*College of Civil Engineering, Department of Geotechnical Engineering, Tongji University, Shanghai, China.*

**Abstract**

Transition metal carbides include a wide variety of materials with attractive properties that are suitable for numerous and diverse applications. Most recent experimental advance could provide a path toward successful synthesis of large-area and high-quality ultrathin Mo$_2$C membranes with superconducting properties. In the present study, we used first-principles density functional theory calculations to explore the mechanical and optical response of single-layer and free-standing Mo$_2$C. Uniaxial tensile simulations along the armchair and zigzag directions were conducted and we found that while the elastic properties are close along various loading directions, nonlinear regimes in stress-strain curves are considerably different. We found that Mo$_2$C sheets present negative Poisson's ratio and thus can be categorized as an auxetic material. Our simulations also reveal that Mo$_2$C films retain their metallic electronic characteristic upon the uniaxial loading. We found that for Mo$_2$C nanomembranes the dielectric function becomes anisotropic along in-plane and out-of plane directions. Our findings can be useful for the practical application of Mo$_2$C sheets in nanodevices.

*Corresponding author (Bohayra Mortazavi): bohayra.mortazavi@gmail.com

Tel: +49 157 8037 8770; Fax: +49 364 358 4511; [#]Timon.rabczuk@uni-weimar.de

## 1. Introduction

The outstanding properties of two-dimensional (2D) materials including graphene [1–4], result in tremendous attention for their usage in a wide variety of intriguing applications such as, nanoelectronics and optoelectronics. During the last decade a huge amount of scientific efforts has been devoted to analyse and synthesize various large-area and high-quality 2D crystals to explore new physics and outstanding properties for advanced practical applications. As a results of these endeavours, currently there exist well-established knowledge for the synthesis of a broad family of 2D materials including: hexagonal boron-nitride [5,6], graphitic carbon nitride [7–9] silicene [10,11], germanene [12], stanene [13], transition metal dichalcogenides such as $MoS_2$ and $WS_2$ [14–16], phosphorene [17,18] and most recently borophene [19] nanomembranes. It is worthwhile to note that various 2D materials can be integrated to form heterostructures [20,21] which can further tune desirable material characteristics for specific applications [22–24]. In line of continuous advances for the synthesis of 2D materials, recently, large-area and high-quality ultrathin $Mo_2C$ membranes have been successfully synthesized by using the chemical vapour deposition (CVD) technique on Cu/Mo foil [25].

$Mo_2C$ exhibits several outstanding properties such as; excellent catalytic activity comparable to the noble metals [25,26], superconductivity [27–29] and high capacity as an anode material to be employed in Li or Na-ion batteries [30], with low diffusion barriers. Motivated by the most recent experimental advance in the fabrication of $Mo_2C$ sheets [25] and their interesting potential applications, various properties of $Mo_2C$ films such as their mechanical, electrical, optical and thermal properties need to be studied to provide a more accurate viewpoint with respect to the application aspects. To this aim, theoretical studies can be considered as a viable approach to evaluate the properties of 2D materials which are complicated, time consuming and expensive to be explored experimentally [31–37]. In particular, mechanical properties of materials play crucial roles since they are strongly related to the structural safety and stability under various applied forces existing during the service of the materials. In the present work, we studied the mechanical and optical properties of single-layer and free-standing $Mo_2C$ sheet using first-principles density functional theory (DFT) method. In particular, we investigated the effect of loading direction on the uniaxial stress-strain response, optical properties and electronic density of states.

## 2. Methods

Density functional theory calculations were performed using the Vienna ab initio simulation package (VASP) [38,39] along with the Perdew-Burke-Ernzerhof (PBE) generalized gradient approximation exchange-correlation functional [40]. We employed projector augmented wave method [41] with an energy cutoff of 450 eV. Super-cells consisting of 36 and 48 atoms were generated to model tensile loading along the armchair and zigzag directions, respectively. However, worthy to mention that since the dynamical effects such as temperature were neglected and the periodic boundary conditions were applied in planar directions, only a unit-cell modelling with 6 atoms was accurate enough to report the mechanical properties of $Mo_2C$. To simulate the mechanical properties of large-area and single-layer $Mo_2C$, one has to apply the periodic boundary condition in planar directions, otherwise the free atoms on the edge may affect the predicted mechanical properties. Periodic boundary conditions were therefore applied in all directions, and a vacuum layer of 20 Å was considered to avoid image-image interactions along the sheet thickness. We used 5×5×1 and 4×4×1 Monkhorst-Pack [42] k-point mesh size for the samples elongated along the armchair and zigzag directions, respectively.

Fig.1, illustrates the $Mo_2C$ structure which has a hexagonal lattice with atomic stacking sequence of ($Mo_{top}$–C–$Mo_{Bot.}$) ABC, in which the Mo atoms on the bottom are placed in the hollow center of the hexagonal lattice. As shown in Fig. 1, the atomic structure of $Mo_2C$ can be well-defined based on the hexagonal lattice constant of (α) and Mo–C bond length. After the energy minimization and geometry optimization using the DFT simulations, the lattice constant and Mo–C bond length were obtained to be 3Å and 2.09 Å, respectively, which are in excellent agreements with the values (2.99 Å and 2.08 Å, respectively) obtained by a recent investigation by Cakir *et al.* [30]. To evaluate the mechanical properties, we elongated the periodic simulation box size in a step-by-step manner in which at every step a small engineering strain of 0.001 was applied. After changing the simulation box size, we rescaled the atomic positions such that no void is formed in the atomic lattice [23,43]. Moreover, at every step of the box size change, very small random displacements along the planar directions were added to the atomic positions to avoid error in the VASP calculation due to the symmetrical atomic positions [23]. After applying the loading conditions, we achieved the energy minimization to update the atomic positions and relax the structure to the ground-state through using the quasi-Newton algorithm as it is implemented in VASP with $10^{-4}$ eV criteria for energy the convergence. After the energy minimization, the stress tensor at each step was acquired to finally report the stress-strain responses. After obtaining the minimized structure, to evaluate the electronic density of states (DOS), we performed a single-point calculation with an energy cutoff of 500 eV in which the Brillouin zone was sampled using a 11×11×1 k-point mesh size. We also used VESTA [44] package for the illustration of the structures.

Optical calculations in this paper were performed in the random phase approximation (RPA) using full potential Wien2k code [45]. Because of metallic properties of $Mo_2C$ sheet, the intraband transitions contribution should be added to the interband transitions to investigate the optical properties of this system. In this case, we used Wien2k package because it provides the possibility to add the intraband transitions contribution to the interband transitions [46,47]. Maximum angular momentum of the atomic orbital basis functions was set to $l_{max} = 10$. In order to achieve energy eigenvalues convergence, wave functional in the interstitial region were expanded in terms of plane waves with a cut-off parameter of $RMT \times K_{max} = 8.5$, where RMT denotes the smallest atomic sphere radius and $K_{max}$ largest k vector in the plane wave expansion. The optical spectra were calculated using 20×20×1 Γ centered Monkhorst-Pack [42] k-point mesh in the first Brillouin zone and setting Lorentzian broadening with gamma equal to 0.05 eV.

The optical properties are determined by the dielectric function $\varepsilon_{\alpha\beta}(\omega) = \text{Re}\,\varepsilon_{\alpha\beta}(\omega) + i\,\text{Im}\,\varepsilon_{\alpha\beta}(\omega)$, which mainly depends on the electronic structure. The imaginary part $\text{Im}\,\varepsilon_{\alpha\beta}(\omega)$ of the dielectric function is obtained from the momentum matrix elements between the occupied and unoccupied virtual wave functions [48,49]:

$$\text{Im}\,\varepsilon_{\alpha\beta}^{(\text{inter})}(\omega) = \frac{4\pi^2 e^2}{\Omega} \lim_{q \to 0} \frac{1}{|q|^2} \sum_{c,v,k} 2 w_k \delta(\varepsilon_{ck} - \varepsilon_{vk} - \omega) \times \langle u_{ck+e_\alpha q} | u_{vk} \rangle \langle u_{ck+e_\beta q} | u_{vk} \rangle^* \quad (1)$$

In this equation $q$ is the Bloch vector of the incident wave and $w_k$ the **k**-point weight. The band indices $c$ and $v$ are restricted to the conduction and the valence band states, respectively. The vectors $e_\alpha$ are the unit vectors for the three Cartesian directions and $\Omega$ is the volume of the unit cell. $u_{ck}$ is the cell periodic part of the orbitals at the $k$-point **k**. The real part $\text{Re}\,\varepsilon_{\alpha\beta}(\omega)$ can be evaluated from $\text{Im}\,\varepsilon_{\alpha\beta}(\omega)$ using the Kramers–Kronig transformation [50]:

$$\text{Re}\,\varepsilon_{\alpha\beta}^{(\text{inter})}(\omega) = 1 + \frac{2}{\pi} P \int_0^\infty \frac{\omega' \text{Im}\,\varepsilon_{\alpha\beta}(\omega')}{(\omega')^2 - \omega^2 + i\eta} d\omega' \quad (2)$$

where $P$ denotes the principle value and $\eta$ is the complex shift. In these two cases, calculations are performed by considering interband transitions. By taking into account the contribution of intraband transitions for metals, it is obtained:

$$\text{Im}\,\varepsilon_{\alpha\beta}^{[\text{intra}]}(\omega) = \frac{\Gamma \omega_{pl,\alpha\beta}^2}{\omega(\omega^2 + \Gamma^2)} \quad (3)$$

$$\text{Re}\,\varepsilon_{\alpha\beta}^{[\text{intra}]}(\omega) = 1 - \frac{\omega_{pl,\alpha\beta}^2}{\omega(\omega^2 + \Gamma^2)} \quad (4)$$

where $\omega_{pl}$ is the plasma frequency and $\Gamma$ is the lifetime broadening.

## 3. Results and discussions

To evaluate the elastic properties of Mo$_2$C films, we first applied uniaxial strains. In Fig. 2, the DFT results for stress-strain response of Mo$_2$C films elongated along the armchair and zigzag directions are plotted. In this case, unidirectional strains are applied along the loading direction while on the transverse direction the box size is remained unchanged. In this case, the stress values along the loading (longitudinal) ($\sigma_l$) and transverse ($\sigma_t$) directions were calculated at each level of loading in order to evaluate elastic properties. The results shown in Fig. 2 present linear relations for the stress values with respect to the strain, which implies that the specimen is stretched within its elastic regime. The Poisson's ratio and elastic modulus were then calculated on the basis of the Hooke's Law. In the general case for a sheet with orthotropic elastic properties, the Hooke's Law has can be written as follows:

$$\begin{bmatrix} s_{xx} \\ s_{yy} \end{bmatrix} = \begin{bmatrix} \frac{1}{E_x} & \frac{-v_{yx}}{E_y} \\ \frac{-v_{xy}}{E_x} & \frac{1}{E_y} \end{bmatrix} \begin{bmatrix} \sigma_{xx} \\ \sigma_{yy} \end{bmatrix} \quad (5)$$

here $s_{ii}$, $\sigma_{ii}$, $v_{ij}$ and $E_i$ are the strain, stress, Poisson's ratio and elastic modulus along the "$i$" direction, respectively. If one applies a uniaxial strain only along the "$x$" direction, it means that the strain is zero along the perpendicular direction $s_{yy}=0$, and therefore one can conclude that:

$$v_{xy} = \frac{\sigma_{yy} E_x}{\sigma_{xx} E_y} \quad (6)$$

However, based on the symmetry of the stress and strain tensors, the following relation exist:

$$\frac{v_{yx}}{v_{xy}} = \frac{E_y}{E_x} \quad (7)$$

By replacing the Eq. 7 in Eq. 6, one can obtain:

$$v_{yx} = \frac{\sigma_{yy}}{\sigma_{xx}} \quad (8)$$

We remind that here, $\sigma_{xx}$ and $\sigma_{yy}$ are equivalent with longitudinal and transverse stresses, respectively. As it is clear, by applying the uniaxial strain, the Poisson's ratio can be obtained based on the stress values. Therefore, through simulating the two uniaxial strain loadings along the armchair and zigzag directions, separately, the elastic modulus and Poisson's ratio values along the both directions can be calculated using the mentioned relations.

For the evaluation of elastic properties, we used low strain levels up to 0.007. In next step, as it is shown in Fig. 2, we fitted straight lines to the DFT data points to calculate the elastic properties. Based on our DFT modelling, the elastic modulus of $Mo_2C$ membranes along the armchair and zigzag directions were obtained to be around 133 GPa.nm and 128 GPa.nm, respectively. For the most of materials with positive Poisson's ratio, when their lattice is stretched in one direction and it is not allowed to relax in the perpendicular direction, it can result in stretching stresses (positive stress values) in the perpendicular (transverse) direction. However, for single-layer $Mo_2C$, our results depicted in Fig. 2 reveal that upon the uniaxial strain loading, the stress in the transverse direction is negative leading to a negative Poisson's ratio. Therefore, free-standing $Mo_2C$ sheet can be categorized as an auxetic material. Our calculations show that the Poisson's ratio of $Mo_2C$ nanomembranes along the armchair and zigzag directions are both around -0.15. Based on our results, we found that the negative Poisson's ratio occurs up to strain levels close to 0.025. We also remind that since we applied periodic boundary conditions, the reported elastic modulus and the Poisson's ratio are size independent.

Next, mechanical properties were evaluated using the uniaxial tensile simulations. To ensure uniaxial stress condition in the sample, after applying the loading strain, the simulation box size along the perpendicular direction of the loading was changed consequently in a way that the transverse stress remained negligible in comparison with stress along longitudinal direction [23]. Acquired uniaxial stress-strain responses of pristine $Mo_2C$ along armchair and zigzag loading directions are illustrated in Fig. 3. As a general trend, first the stress-strain curves present linear trends which is followed by nonlinear relations up to the ultimate tensile strength point at which the material illustrates its maximum load bearing ability. Our results shown in Fig. 3 suggest different anisotropic mechanical response of single-layer $Mo_2C$ when it is stretched along the armchair and zigzag direction. We found that the tensile strength along the zigzag is 13.2 GPa.nm whereas this value is 9.5 GPa.nm along the armchair direction. Moreover, the strain at tensile strength along the zigzag direction is 0.27 which is more than twice of that along the armchair direction (0.125).

Deformation process of single-layer $Mo_2C$ stretched along the armchair and zigzag direction are illustrated in Fig. 4. The structures are depicted at five different strain points with respect to the strain at the ultimate tensile strength, $s_{uts}$. Results shown in Fig. 4 reveal that at strain levels higher than $0.5s_{uts}$, the periodic simulation box size along the transverse direction decreases which implies that the negative Poisson's ratio is valid only at low strain levels. For the both loading directions, upon the stretching the sheet's thickness gradually decreases by increasing the strain level. In this case, at the ultimate tensile strength the $Mo_{top}$–$Mo_{Bot}$ distance decreases by around 22% and 10% for the sample elongated along zigzag and armchair direction, respectively. Upon the uniaxial tensile loading, the bonds that are along the direction of stretching were elongated with increasing strain levels, whereas those bonds that are oriented almost toward the perpendicular direction were contracted slightly. For the sample stretched along the armchair direction, in a unit-cell two Mo–C bonds are oriented close to the perpendicular direction and only one Mo–C bond is exactly along the loading direction. The stretchability of this bond along the loading direction dominates the mechanical response of $Mo_2C$. On the other side, for the $Mo_2C$ sheet elongated along the zigzag, in a unit-cell two bonds are almost oriented along the loading direction and the other bond is exactly along the perpendicular direction. In this case, these two bonds are stretching and therefore involved in the load transfer. Moreover, the bond along the perpendicular direction of the loading contract more severely which helps the structure to extend up to higher strain levels and consequently enhance the stretchability of the structure. The ability of the structure to contract in the perpendicular direction of the loading, plays a significant role on the mechanical response. In another word, the highly anisotropic mechanical response of $Mo_2C$ along armchair and zigzag directions can be attributed to fact that along the zigzag the structure can extend at higher strain levels because more bonds are simultaneously involved in stretching and the structure can contract more along the perpendicular and thickness directions.

In addition, as it is shown in the stress-strain relation along the zigzag direction, interestingly, after the strain level of ~0.1, a yield point is observable. At this point, we found that by applying the further strain levels, the Mo-C bond lengths remained unchanged. In this case, by increasing the strain levels the $Mo_2C$ flow easier by contracting more along the thickness ($Mo_{top}$–$Mo_{Bot}$ distance decreases). The bond lengths remain almost unchanged up to strain level of ~0.15. We note that at strain levels around ~0.125, it was observed that the contraction of the sheet along the thickness slow down and therefore the structure start to stretch harder along the loading direction and that explains the increase in the stress values after this point. After the strain of ~0.15 for the $Mo_2C$ elongated along the zigzag direction, the bonds that are oriented along the loading directions start to stretch again and this way the stress values increase beyond the stress value reached first at the strain level of ~0.1.

The electronic density of states (DOS) of single-layer $Mo_2C$ films at different strain levels stretched along both of the armchair and zigzag directions up to the tensile strength point were calculated. Samples of computed total DOSs for the $Mo_2C$ nanomembranes elongated along the armchair and zigzag directions are illustrated in Fig. 5. According to the DOS results for the relaxed and stretched samples, the DOS is not zero at the zero state energy (Fermi level) which accordingly demonstrates metallic characteric. Based on our first-principles simulations, the total DOS for valence/conduction bands around the Fermi level for the highly strained structures are close to that for the system under no loading condition. We note that we also applied the biaxial loading condition and repeated our DFT modelling. From the total DOS results for the $Mo_2C$ films under biaxial loading and at different strain levels, we found that no bandgap can be opened since all structures present metallic electronic character. Therefore, according to our DFT results, there exist no practical possibility to open a bandgap in free-standing and defect-free $Mo_2C$ by applying uniaxial or biaxial loading conditions.

We note that we also optimized $Mo_2C$ lattice with wien2K code and the lattice constant and Mo–C bond length were obtained to be 2.99Å and 2.08 Å, respectively, which are in excellent agreements with the values obtained by the VASP code and they are exactly the same as those reported in the work by Cakir *et al.* [30]. In addition to further verify that our calculations using Wien2k code are in agreement with those based on VASP code, we also performed the electronic band structure calculations for free of strain $Mo_2C$ using the both packages. Fig. 6, illustrates the electronic band structure of $Mo_2C$ sheet calculated using the VASP and Wien2K cods. It can be seen that the band structure obtained by Wien2k code is in an excellent agreement with that obtained by VASP code and they both confirm the metallic electronic character of the single-layer and free-standing $Mo_2C$.

Fig. 7 and Fig. 8 illustrate Im$\varepsilon$ and Re$\varepsilon$ of strained $Mo_2C$ at different magnitudes of strains for light polarizations parallel (E∥x and E∥y) and perpendicular (E∥z) to the plane within the low-frequency regime (up to 8 eV). In the top panel of each figure, we have reported optical spectra considering the interband transitions contribution. In the below panel of these figures the optical spectra for interband+ intraband transitions contributions has been shown. For all cases, Im$\varepsilon$ obtained from interband transitions contribution in parallel polarizations (for both E∥x and E∥y) starts without a gap, confirms that all aforementioned systems have metallic property. While metallic properties for perpendicular polarization of these systems are weak. Furthermore, there are large sharp peaks in low frequencies around 1 eV for strained systems in parallel polarizations, which were utterly missing in the free-strain case. By adding the intraband transitions contribution, the optical spectra of these systems in the electric field parallel to plane is changed significantly, while the optical spectra in the electric field perpendicular to plane E∥z is not changed which shows semiconductor property. In general, $Mo_2C$ sheet has metallic properties, because the conduction band and valance band cross each other at Fermi level (Fig. 6), it is noteworthy that one peak of DOS for monolayer $Mo_2C$ sheet appears at the Fermi level (Fig. 5). The optical properties of $Mo_2C$ are found to be dependent on the directions of the electric field polarization. $Mo_2C$ sheet in *E//x* presents optically metallic property but in *E//z*, because of the huge depolarization effect, illustrates semiconductor property. Due to the huge depolarization effect in the 2D planar geometry for light polarization perpendicular to the plane [51], the optical properties for light polarization parallel to the plane are very important. These properties have also seen for graphene [49,52,53] and others metallic 2D sheets like $B_2C$ monolayer sheet [54]. It can be seen that the values of static dielectric constant (the values of dielectric constant at zero energy) in parallel polarizations are greater than 19 within interband transitions contribution; these values in E∥z are at range of 7-8.5. By taking into account the intraband transitions contribution, the Re$\varepsilon$ for all cases in E∥x and E∥y is changed significantly, while it does not have any effect on the Re$\varepsilon$ in perpendicular polarization. The roots of real part of dielectric function with x = 0 line show the plasma frequencies [54]. Table 1 summarizes our results of the first plasma frequency for $Mo_2C$ sheet at different magnitudes of strain for both parallel and perpendicular polarizations. The first plasma frequency for all systems in E∥z is nearly zero which confirms theses materials have semiconductor properties in perpendicular polarization. Furthermore, by considering the intraband transitions contribution, the imaginary and real parts of the dielectric function have a singularity at zero frequency, as can be seen in Figs. 7 and Fig. 8. Our results show that for all aforementioned systems the dielectric function becomes anisotropic along armchair, zigzag and out-of plane directions. The absorption coefficient is calculated using the following relation

$$a_{\alpha\beta}(\omega) = \frac{2\omega k_{\alpha\beta}(\omega)}{c} \quad (9)$$

where $c$ is the speed of light in vacuum and $k_{\alpha\beta}$ is imaginary part of the complex refractive index, known as the extinction index. It is given by the following relations

$$k_{\alpha\beta}(\omega) = \sqrt{\frac{|\varepsilon_{\alpha\beta}(\omega) - \operatorname{Re}\varepsilon_{\alpha\beta}(\omega)|}{2}} \quad (10)$$

For all cases, absorption coefficient in all directions of electric field is shown in Fig. 9. It is found that there are two main peaks in all polarizations. First main peak occurs at energy around eV that is related to π electron plasmon peak. Second main peak occurs around 35–40 eV that is related to π+σ electron plasmon peak. At the energies range of 0-20 eV value of absorption in E∥x and E∥y is more than value of absorption in E∥z. Moreover, at the energies range of 20-35 eV, the value of absorption for all cases is very small.

## 4. Conclusion

In summary, first-principles density functional theory simulations were carried out to explore the mechanical and optical response of free-standing and single-layer $Mo_2C$. We predicted elastic modulus of 133 GPa.nm and 128 GPa.nm along the armchair and zigzag directions, respectively. It is found that $Mo_2C$ is an auxetic material with negative Poisson's ratio. The Poisson's ratios of single-layer $Mo_2C$ along the armchair and zigzag directions are both around -0.15. According to our modelling results, $Mo_2C$ presents anisotropic mechanical response with distinct strain-stress curves. It is worth noting that $Mo_2C$ nanomembranes can exhibit high tensile strength of 13.2 GPa.nm and 9.5 GPa.nm along the zigzag and armchair directions, respectively. We found that the strain at tensile strength along the zigzag direction is around 0.27 which is more than twice of that along the armchair direction (0.125). Considerable anisotropic mechanical response of Mo$_2$C along the armchair and zigzag direction were attributed to the fact that along the zigzag the structure can extend more extensively due to the presence more bonds involved in the stretching and the structure may contract more along the perpendicular and thickness directions. The electronic density of states calculations of the relaxed and highly stretched Mo$_2$C structures, reveal the metallic characteristic for the all structures and propose that one cannot open a bandgap in Mo$_2$C through applying uniaxial or biaxial mechanical loading conditions. By adding the intraband transitions contribution, the optical spectra of relaxed and stretched systems in the electric field parallel to plane is changed significantly, while the optical spectra in the electric field perpendicular to plane E||z is not changed which shows semiconductor property. Our results show that for all aforementioned systems the dielectric function becomes anisotropic along armchair, zigzag and out-of plane directions. Our investigation proposes the Mo$_2$C nanomembranes as an auxetic material with considerable stretchability and mechanical strength.

**Acknowledgment**

BM and TR greatly acknowledge the financial support by European Research Council for COMBAT project (Grant number 615132).

Table 1. Calculated the first plasma frequency at different magnitudes of strain for both parallel and perpendicular polarizations.

| Systems | E∥x | E∥y | E∥z |
|---|---|---|---|
| $s=0$ | 2.46 | 2.60 | 0.09 |
| $s=0.5\ s_{uts}$ Armchair | 2.32 | 2.70 | 0.08 |
| $s=s_{uts}$ Armchair | 2.26 | 3.37 | 0.08 |
| $s=0.5\ s_{uts}$ Zigzag | 3.07 | 2.63 | 0.07 |
| $s=s_{uts}$ Zigzag | 2.89 | 3.00 | 0.07 |

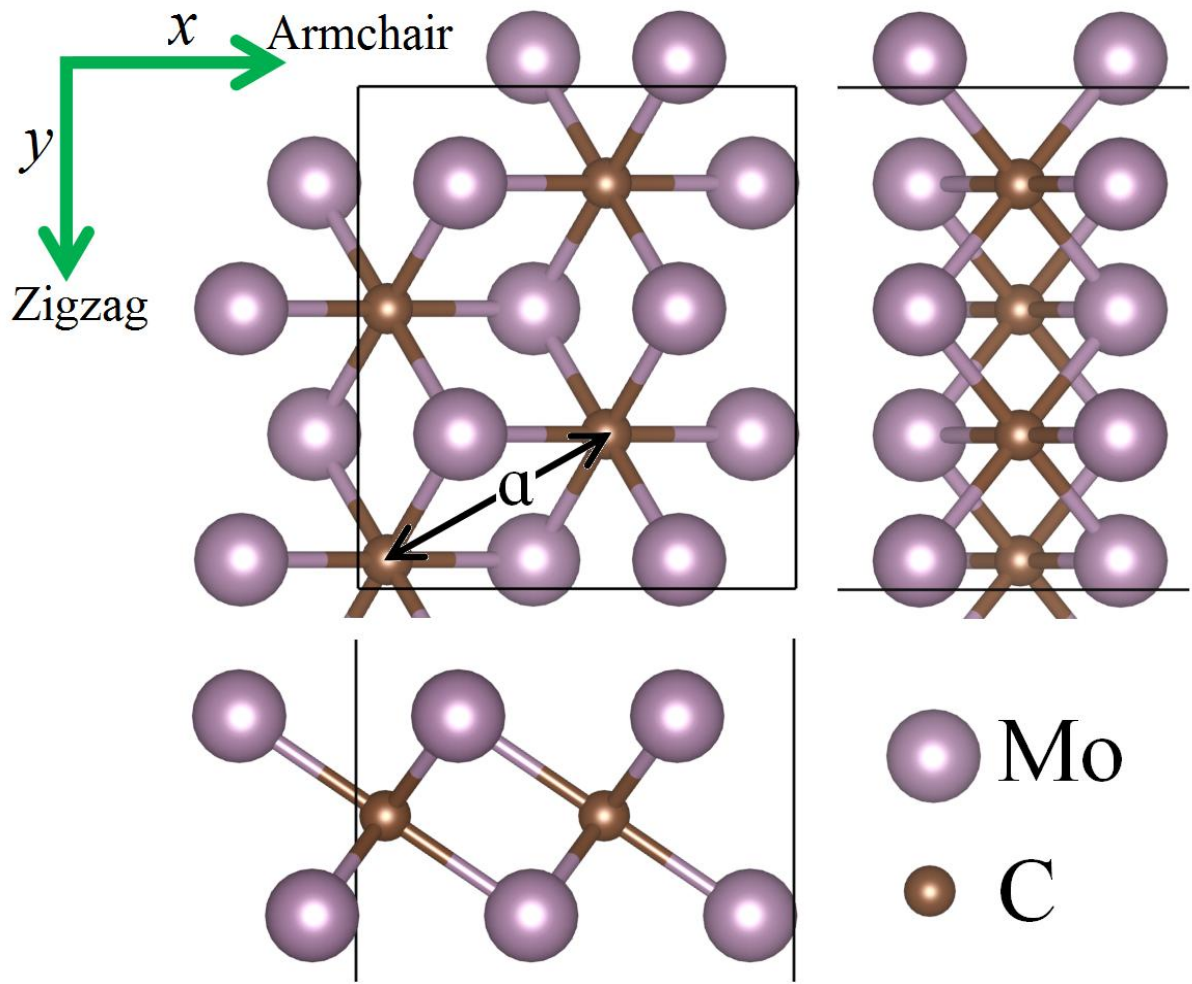

Fig. 1, Top and side views of atomic configuration in Mo$_2$C lattice. Mechanical properties are studied along the armchair and zigzag directions as shown here.

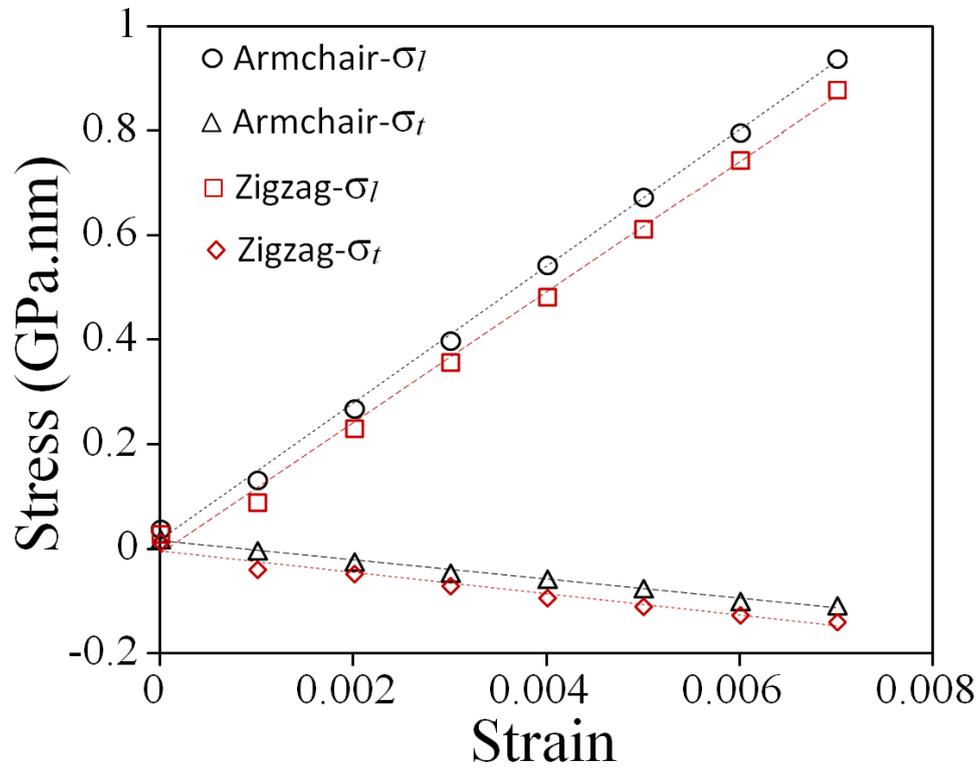

Fig. 2, First principles density functional theory results of stress-strain response of Mo$_2$C films stretched along the armchair and zigzag directions while the box size in the perpendicular direction was fixed. Here, $\sigma_l$ and $\sigma_t$ denote the stress values along the longitudinal (loading) and transverse (perpendicular of loading) directions, respectively.

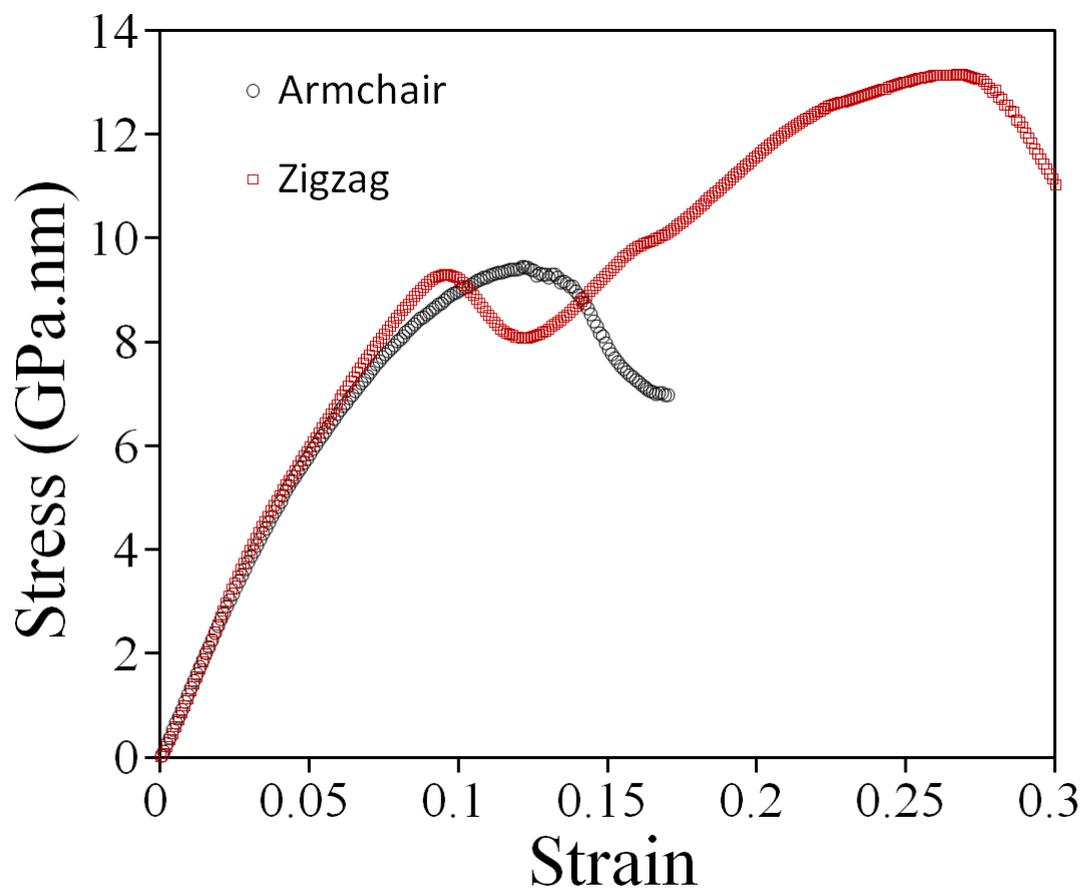

Fig. 3, Uniaxial stress-strain response of defect-free and free-standing single-layer $Mo_2C$ elongated along the armchair and zigzag directions.

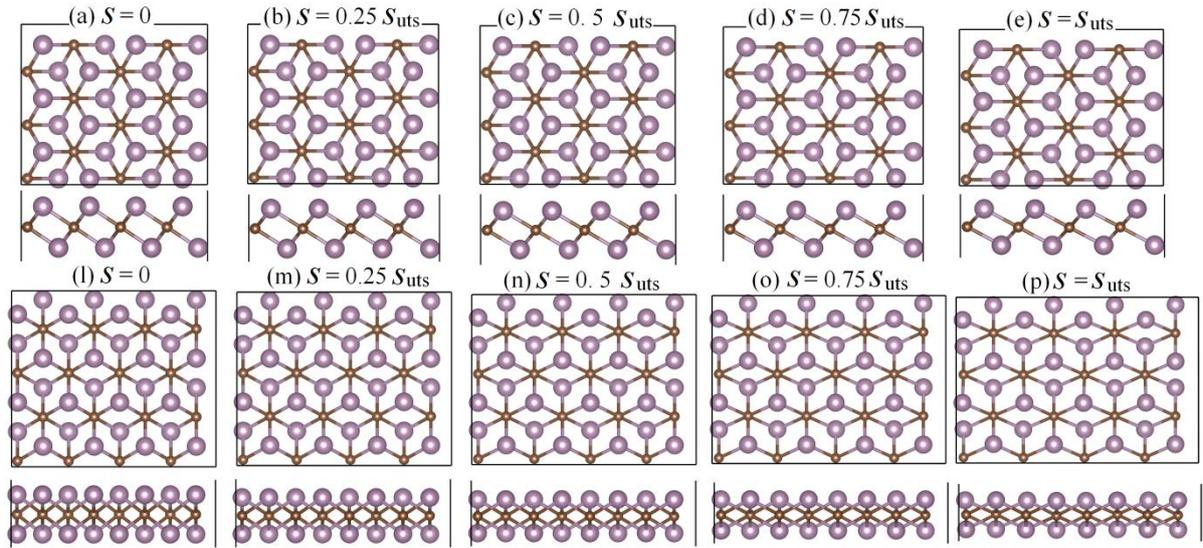

Fig. 4, Top and side view of uniaxial tensile deformation processes of single-layer $Mo_2C$ elongated along the armchair (a-e) and zigzag (l-p) at different strain levels ($s$) with respect to the strain at ultimate tensile strength ($s_{uts}$).

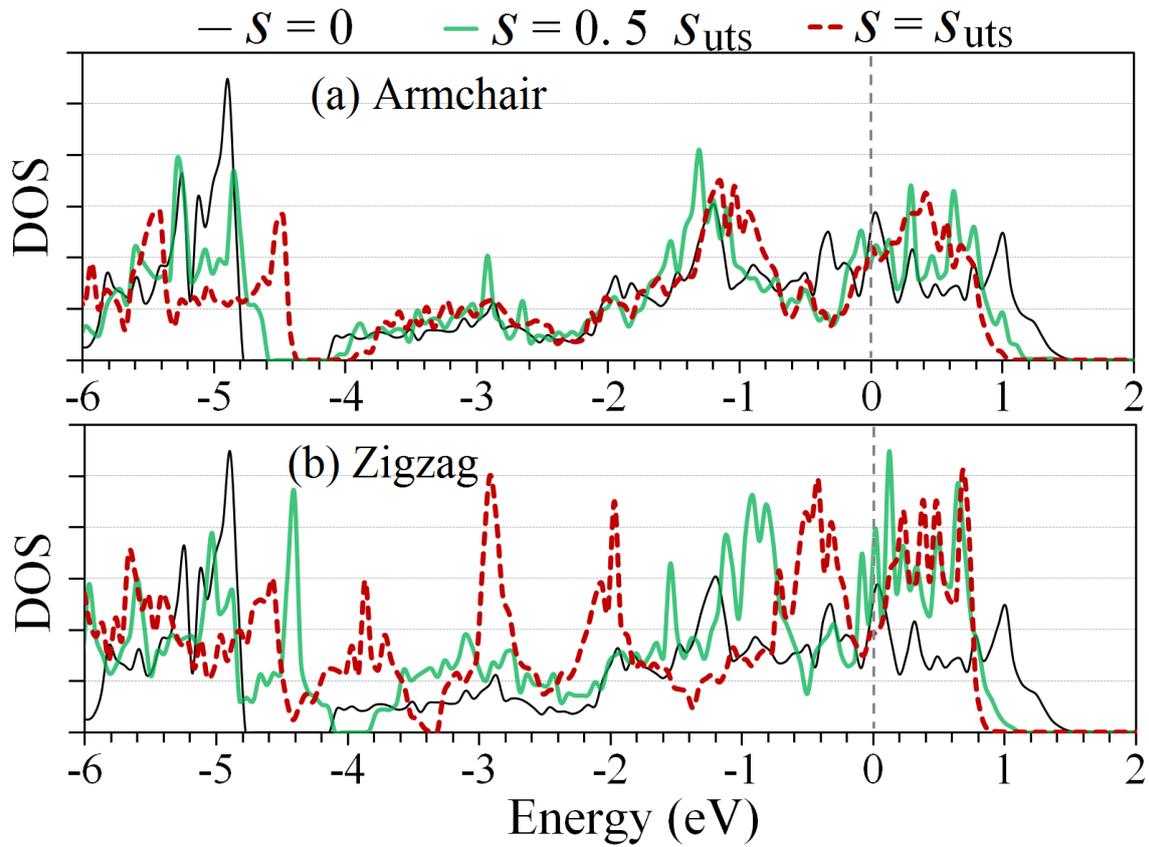

Fig. 5, Electronic density of states (DOS) for uniaxially loaded $Mo_2C$ along the (a) armchair and (b) zigzag directions. For each case, the results are depicted for three strain levels, $s$, with respect to the strain at ultimate tensile strength ($s_{uts}$).

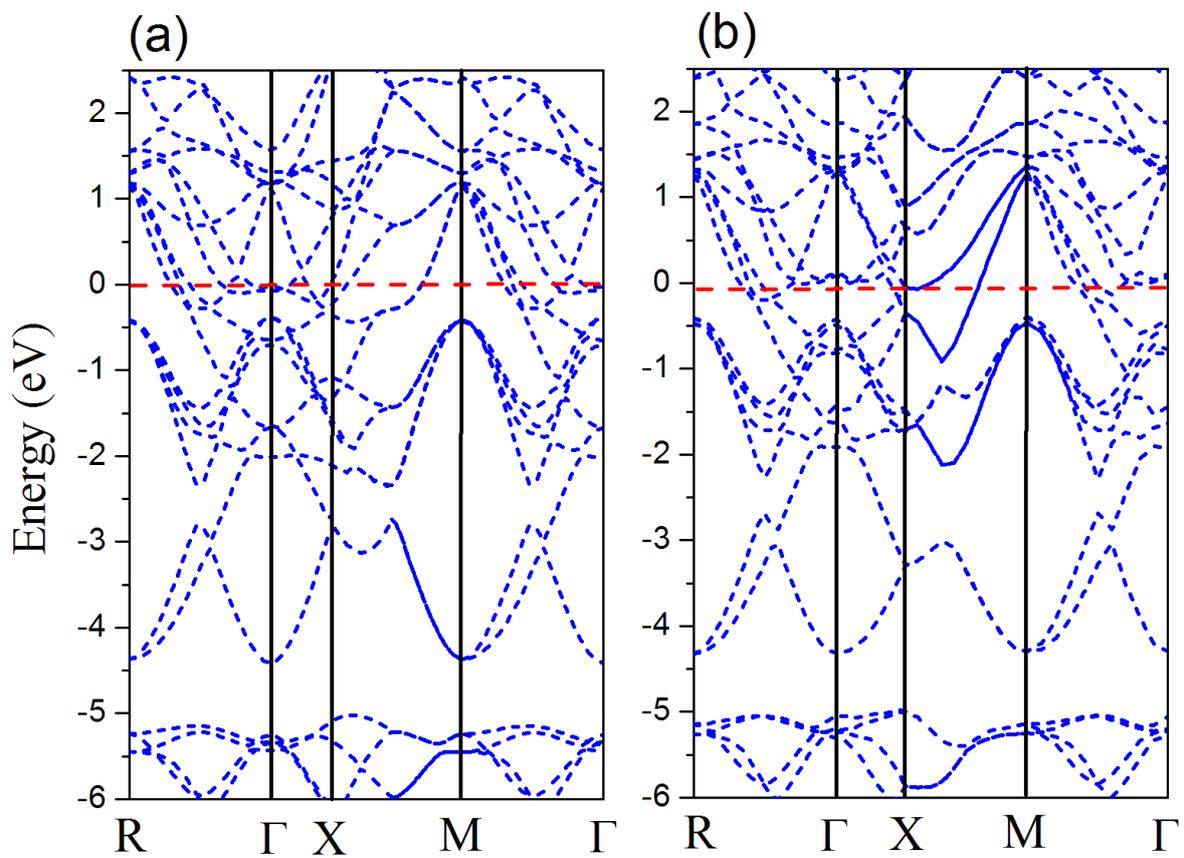

Fig. 6, Electronic band structure of stress-free and single-layer $Mo_2C$ sheet calculated with (a) VASP and (b) Wien2k.

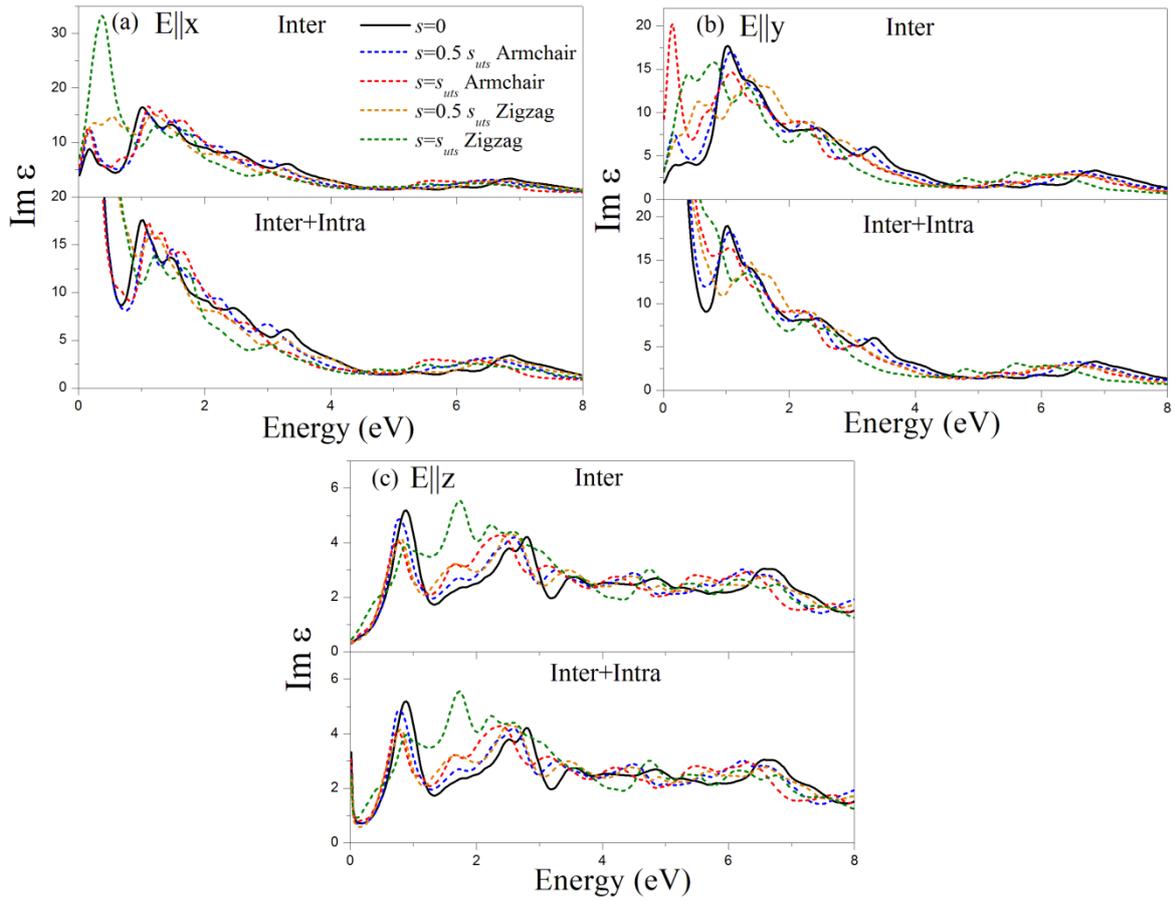

Fig. 7, Imaginary part of the dielectric function of strained $Mo_2C$ at different magnitudes of strains for light polarizations (a) parallel to the x-axis (E||x) (b) parallel to the y-axis (E||y) and parallel to the z-axis (E||z). Below panel in each figure illustrates Im$\varepsilon$ considering the intraband contribution in addition to the interband one.

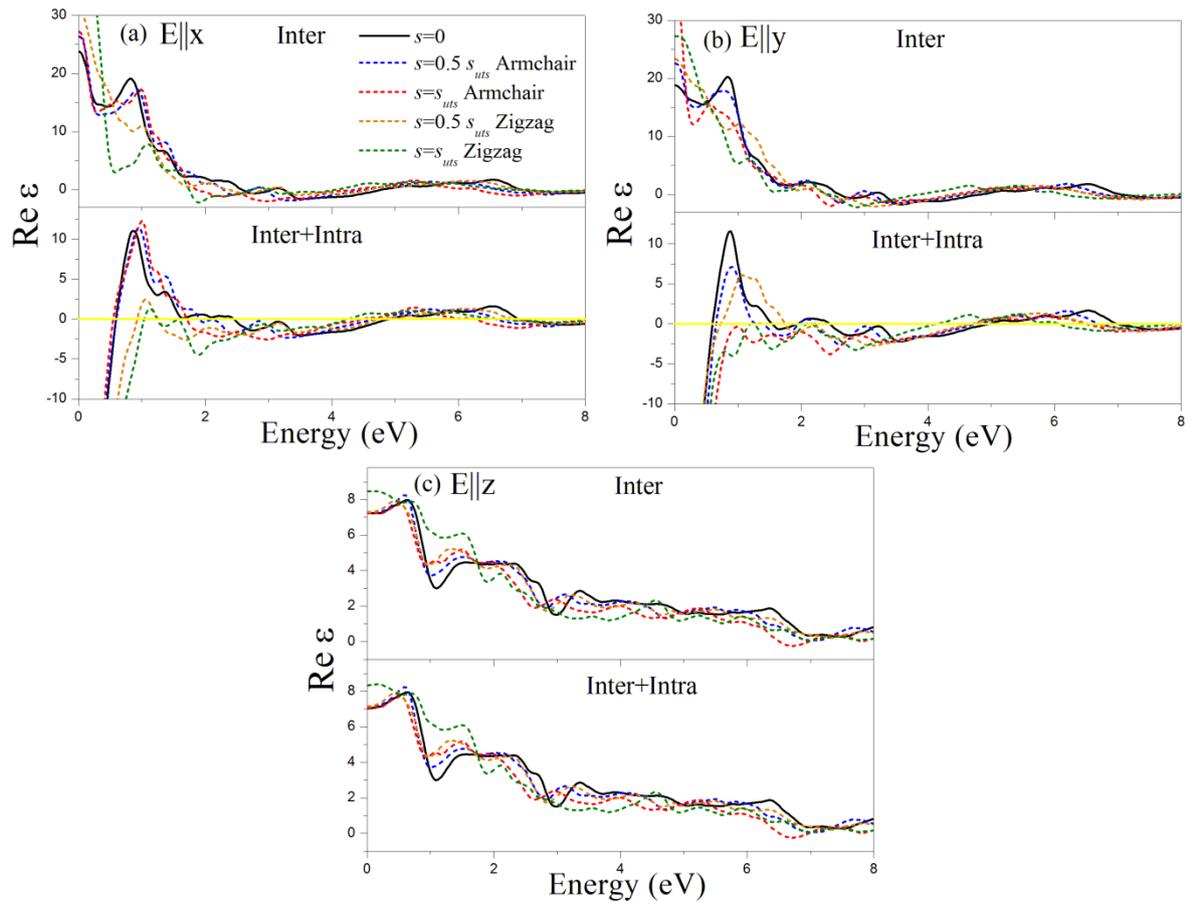

Fig. 8, Real part of the dielectric function of strained $Mo_2C$ at different magnitudes of strains for light polarizations (a) parallel to the x-axis (E||x) (b) parallel to the y-axis (E||y) and parallel to the z-axis (E||z). Below panel in each figure illustrates Re$\varepsilon$ considering the intraband contribution in addition to the interband one.

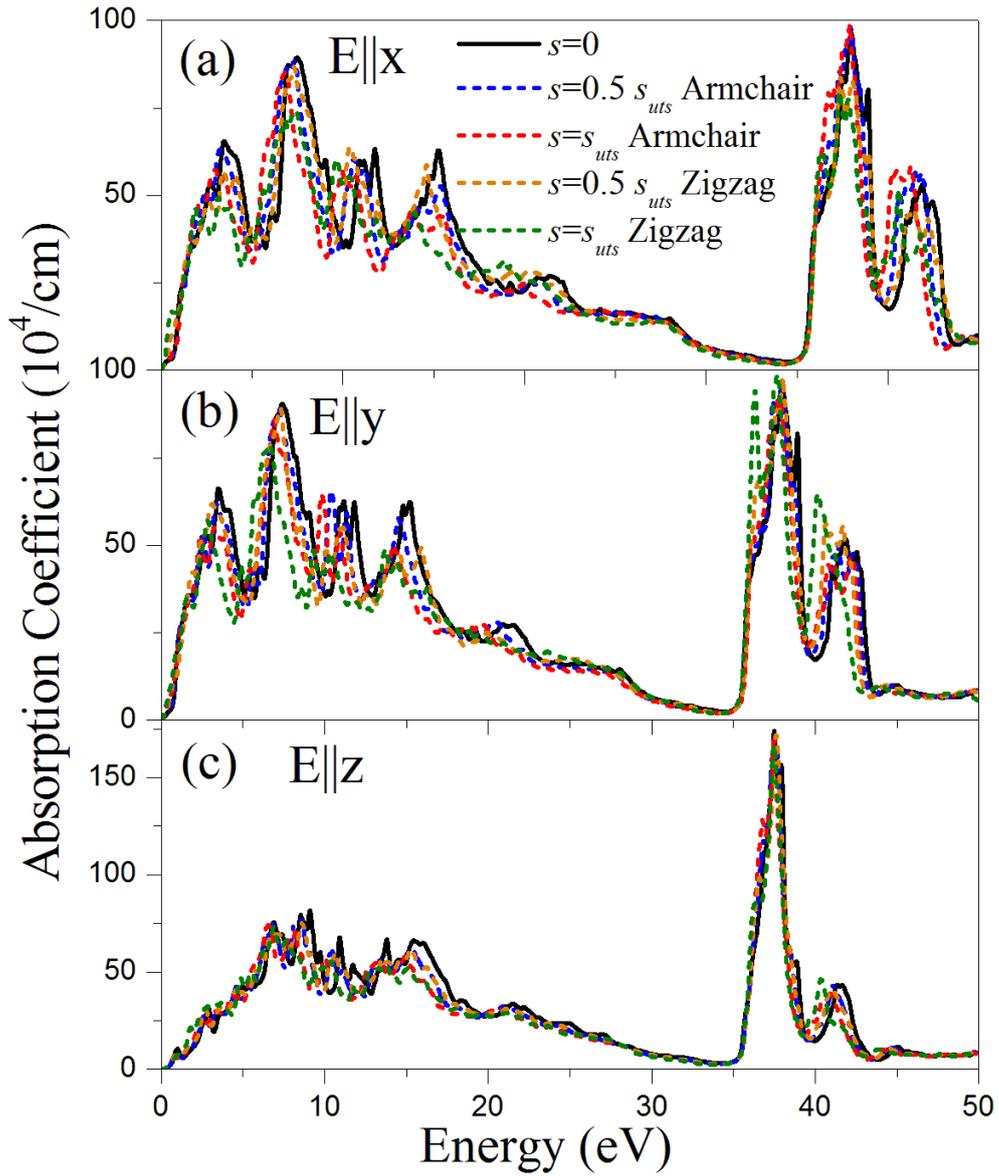

Fig. 9, The absorption coefficient of strained $Mo_2C$ at different magnitudes of strains for light polarizations (a) parallel to the x-axis (E∥x) (b) parallel to the y-axis (E∥y) and parallel to the z-axis (E∥z).